# Definition and Research of Internet Neurology

Feng Liu

School of Computer and Information Technology, Beijing Jiaotong University, Beijing 100044, China
Email: zkyliufeng@126.com

**Abstract**

More and more scientific research shows that there is a close correlation between the Internet and brain science. This paper presents the idea of establishing the Internet neurology, which means to make a cross-contrast between the two in terms of physiology and psychology, so that a complete infrastructure system of the Internet is established, predicting the development trend of the Internet in the future as well as the brain structure and operation mechanism, and providing theoretical support for the generation principle of intelligence, cognition and emotion. It also proposes the viewpoint that the Internet can be divided into Internet neurophysiology, Internet neuropsychology, Brain Internet physiology, Brain Internet psychology and the Internet in cognitive science.

## 1. Research Background of Internet Neurology

Every major technological change in the human society will lead to a scientific revolution in a new area. The biological diversity and the influence of isolated ecosystems on organisms were witnessed by human beings in the Uncharted Waters. However, Either Darwin or Wallace found the biological evolution phenomenon by following the fleet in voyage.

Large Industrial Revolution enabled human had a great improvement either in the use of power or observation ability and had technologically supported major physics breakthrough starting 100 years later, such as Newton's Universal Gravitation, Einstein's Theory of Relativity and the quantum mechanics building created by many scientists. All of these breakthroughs relate to the "force" and "observation".

The influence of the Internet revolution on human beings has far exceeded that of the Great Industrial Revolution. Different from the Industrial Revolution which enhanced the human strength and extended their vision, the Internet greatly has promoted human intelligence, enriched human's knowledge, while intelligence and knowledge are just the factors the most closed to the brain.

## 2. Preliminary Theoretical Work about Internet Neurology

On September 23, 2008, the author of this paper, along with Peng Qiu and Liu Ying from University of Chinese Academy of Sciences, issued an article with the title of Discovery and Analysis of Internet Evolution Law, where, it was presented that "the Internet and the human brain are very complementary to the scientific research. We are likely to have chances to understand characteristics of both at the same time in the next few decades. Based on the known structure of human brain, we can predict the development trend of the Internet in the next step (note that it's not planning). Similarly, with the exceptionally fast evolution of the Internet, we will be able to find more secrets of human brain via contrast. Now, it may be asserted that there should be an address coding system and a search engine system in the human brain. We are looking forward to neuroscientists' new discovery in this regard." **[1]**. An Internet-like brain structure diagram was given in that article, as shown in Figure 1:

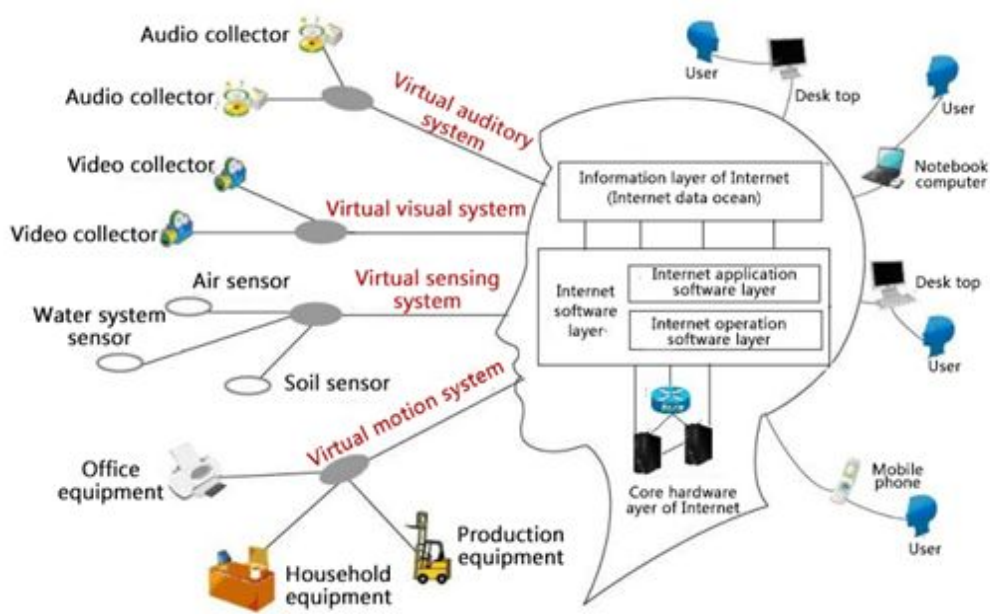

Fig. 1 Structure Diagram of the Internet Virtual Brain

In 2010, the author of this paper issued another article with the title of Cross-Contrast Research of the Internet and Brain Science in the journal of Complex Systems and Complexity Science. The article describes the structure and functions of Internet virtual brain from aspects of the virtual neurons, virtual sensory and motor systems, the virtual autonomic nervous system, the virtual central nervous system, and the virtual nerve reflex arc, based on the inspiration of neurology on the Internet research; and, introduces research methods and the experimental design regarding functions of the Internet-like structure in the human brain in terms of the routing system, search engine application, Wikipedia application and IP address application, based on the inspiration of the Internet on the neurological research **[2]**.

In March 2010, an experiment on the human brain including the search engine function was carried out by the author of this paper in one of classrooms of University of Chinese Academy of Sciences. The experiment results were published in the book titled as Internet Evolution Law issued in September 2012. This book introduces the composition of each Internet nervous system, deeply discusses the operation mechanism of Internet nerve reflex arc, and describes Internet-like phenomena in the human brain. Besides, the description about the social networking-like features in the brain is involved in the book, and the contents are as follows: There should be short message communication like microblogging (SNS) between brain neurons. Accordingly, the functions in microblogging like adding friends or following should also exist. That means the information transmission occurs between different brain neurons. Whether the human brain contains an operation mechanism similar to microblogging has not been proved, but viewing from the actual brain response characteristics relative to information, we believe it is a scientific conjecture possible to be proved. As long as neural scientists focus on analyzing and summarizing the operation characteristics of microblogging, and design experimental methods based on which, this conjecture may eventually be proved in the future [3].

From 2012, scientific researchers in United States and other countries also began to notice the correlation between the Internet and brain science. On November 16, 2012, Dmitri Krioukov, from University of California San Diego, issued an article with the title of Network Cosmology in Scientific Report, where, he presents that there is a high similarity between the Internet and the brain neural network in terms of development and construction [4].

On February 4, 2015, the researchers from the University of Basel reported that "it has been found that the neurons in human brain are linked with each other, like a social network. Each nerve cell is linked with many other cells, but only a few ones that are very similar to each other have a strong link". These findings were published in the magazine Nature on February 4, 2015 [5].

## 3. Presentation and Composition of Internet Neurology

Internet and neurology are originally the fields distant to each. Nowadays, however, they seem to have a deep and close relation. According to relevant theories and practice foundations built in the past 7 years, the crossover of the two fields produces a new discipline in the 21st century, i.e. Internet neurology, which may be defined like this:

Based on neurological research achievements, the Internet hardware structure, software system, data and information and commercial applications may be organically integrated to form a complete structure system of Internet, and predict the new functions and architectures that may be generated from the Internet along the path of neurology; to propose imagines and analyze the biological foundation for the generation of consciousness, thinking, intelligence and cognition in human brain, based on the continuously produced and stabilized structure functions; and, to research how the Internet and the human brain-two giant systems- form a cross relation of mutual influencing and modeling as well as combining and evolving with each other.

If a coordinate system is established with brain science and the Internet respectively as the two sections of X-axis on both sides, and physiology and psychology as the two sections of Y-axis, then the Internet neurology consists of four parts, i.e. Internet neurophysiology, Internet neuropsychology, Brain Internet physiology and Brain Internet psychology. Besides, the crossover part of the above four forms the fifth member-the Internet in cognitive science. Their relationship is as shown in Fig.2.

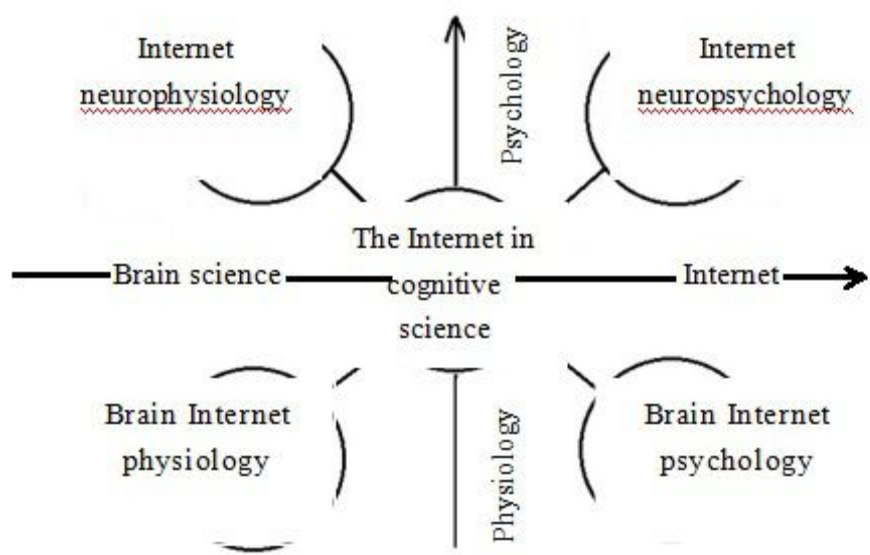

Fig.2 Composition of Internet neurology

## 4. Brief Introduction of Five Disciplines in Internet Neurology

### 4.1 Internet Neurophysiology

It mainly involves the research on neurology-based basic functions and structures of the Internet, including but not limited to, Internet central nervous system, Internet sensory nervous system, Internet sports nervous system, Internet autonomic nervous system, Internet reflex arc and algorithm deep-learning basis; and artificial intelligence processing mechanism including image, voice and video identification by applying Internet big data.

### 4.2 Internet Neuropsychology

It mainly involves the research on Approximatively-neuropsychological Internet phenomena generated in the process that the Internet is evolving towards the mature brain, including but not limited to, generation of Internet collective intelligence,

Internet emotion, generation and characteristics of Internet dreams as well as Internet IQ.

**4.3 Brain Internet Physiology**

It mainly involves the research on the conditions similar to Internet structure functions in the human brain, so that the Internet will serve as a reference system for people to interpret the biological principle of the human brain, including but not limited to the search engine mechanism, the Internet-like routing mechanism, the IPv4 / IPv6-like mechanism, the neuron-like social network interaction mechanism and the influence of human using the Internet on the remodeling of brain physiological structure.

**4.4 Brain Internet Psychology**

It mainly involves the research on the influence and remodeling on the human brain in terms of psychology applied by the Internet, including but not limited to, users' habitual craving and influence on their intelligence, emotion and social relationship induces by the Internet.

**4.5 The Internet in Cognitive Science**

It can be seen as the cross part of Internet neurophysiology, Internet neuropsychology, brain Internet physiology and brain Internet psychology and mainly involves the research on the deep principle for that the two giant systems i.e. the Internet and the human brain generate intelligence, cognition and emotion as the results of mutual influencing and modeling as well as combining and evolving with each other.

**5. Subsequence Research**

Internet Neurology is a new scientific theory forming based on the research of the past ten years, with the goal to establish a complete Internet architecture through the cross-contrast between the Internet and the brain science in term of physiology and psychology, so that people can predict the development trend of Internet in the future as well as the structure and operation mechanism of brain, and provide theoretical support for the generation principle of intelligence, cognition and emotion.

16 questions are presented as the research focus of further Internet neurology. The questions respectively belong to Internet neurophysiology, Internet neuropsychology, brain Internet physiology, brain Internet psychology and the Internet in Cognitive Science, as given in the table below.

**Table 1 The Sixteen Questions for Research of Internet Neurology**

| Discipline | Question |
| --- | --- |
| Internet neurophysi | 1. Is the Internet evolving towards the state with high a similarity relative to the brain functions and structure? Is there any rule to follow? |
| ology | 2. If the Internet forms a brain-like structure, then, what is the composition of this structure? And what is its operation mechanism? |
| | 3. What inspirations are obtained from neurology in the evolution process of Internet? Then, what new valuable functions and applications are predicted to appear? |
| Internet neuropsychology | 1. Will the Internet brain have psychological phenomena as the human brain? |
| | 2. How does the dream system of the Internet brain consist? And how is it helpful to interpret the human dream? |
| | 3. Can the Internet brain perform the IQ determination? And, is it helpful to identify the relationship between human intelligence and machine intelligence? |
| | 4. How are the intelligence and emotion of the Internet brain generated? And, what influence will they have on the real human social? |
| Brain Internet physiology | 1. Is there any application and function similar to the Internet in the human brain? How do such typical Internet applications as SNS, search engine, routing, IP, etc. exist in the brain? |
| | 2. When were the Internet-like applications and functions evolved? And, how will they evolve in the future? |
| | 3. What influence does the Internet have on the development and evolution of the human brain? |
| Brain Internet psychology | 1. What influence will the Internet have on the human's social psychology, with its constant involvement in the human's life? |
| | 2. Is the formation of habitual craving on Internet an inevitable aspect generated from the symbiotic evolution of the Internet and human? |
| | 3. What influence will the Internet have on the human intelligence? |
| The Internet in cognitive science. | 1. How do the Internet and the brain mutually influence and model? |
| | 2. What is the deep reason for that the Internet evolves towards the brain, and the brain contains some Internet-like applications? |
| | 3. What influence will the cross-contrast between the Internet and the human brain have on the research about the generation of intelligence, cognition and emotion? |

## Introduction of authors:

**Liu Feng, male, Dr. M**

School of Computer and Information Technology, Beijing Jiaotong University, Beijing 100044, China

**Email: zkyliufeng@126.com;**

Research Areas:

Interdisciplinary study of the Internet and the brain science, and study of Internet IQ and artificial intelligence.